\documentclass[12pt,preprint]{aastex}

\usepackage{natbib}
\def\degree{\ifmmode {^\circ}\else {$^\circ$}\fi}
\def\rstar{\ifmmode {\, R_{\star}}\else $R_{\star}$\fi}
\def\msol{\ifmmode {\, M_{\odot}}\else $M_{\odot}$\fi}
\def\rsol{\ifmmode {\, R_{\odot}}\else $R_{\odot}$\fi}
\def\lsol{\ifmmode {\, L_{\odot}}\else $L_{\odot}$\fi}
\def\msolyr{\ifmmode {\, M_{\odot}\,{\rm yr}^{-1}}\else $M_{\odot}\,{\rm yr}^{-1}$\fi}
\def\mdot{\ifmmode {\,\dot{M}}\else $\dot{M}$\fi}
\def\mdotyr{\ifmmode {\,\dot{M}\,yr^{-1}}\else $\dot{M}\,yr^{-1}$\fi}

\newcommand{\Vinf}{\mbox{$V_{\infty}$}~}
\newcommand{\kms}{km s$^{-1}$}
\newcommand{\Teff}{\ifmmode{T_{\rm eff}}\else{$T_{\rm eff}$}}

\begin{document}

\title{Variable Winds and Dust Formation in R Coronae Borealis Stars}

\author{ Geoffrey C. Clayton\altaffilmark{1}, T.R. Geballe\altaffilmark{2}, Wanshu Zhang\altaffilmark{1}}

\altaffiltext{1}{Department of Physics \& Astronomy, Louisiana State
University, Baton Rouge, LA 70803; gclayton,@fenway.phys.lsu.edu, wzhan21@lsu.edu}

\altaffiltext{2}{Gemini Observatory, 670 N. A'ohoku Place, Hilo, HI 96720;  
tgeballe@gemini.edu}






\begin{abstract}

We have observed P-Cygni and asymmetric, blue-shifted absorption profiles 
in the He I $\lambda$10830 lines of twelve R Coronae Borealis (RCB) 
stars over short (1 month) and long (3 year) timescales to look for variations linked to their dust-formation episodes.
In almost all 
cases, the strengths and terminal velocities of the line vary significantly and
are correlated with dust formation events. Strong absorption features 
with blue-shifted velocities $\sim$400 \kms~appear during 
declines in visible brightness and persist for about 100 days after recovery to maximum brightness. Small residual winds of somewhat lower velocity are 
present outside of the decline and recovery periods. The correlations 
support models in which recently formed dust near the star is propelled 
outward at high speed by radiation pressure and drags the gas along with it.

\end{abstract}



\section{Introduction}

The R Coronae Borealis (RCB) stars are a small group of carbon-rich, hydrogen-deficient, post-AGB
supergiants. Almost 100 RCB stars are known in the Galaxy
and the Magellanic Clouds \citep{1996PASP..108..225C,2012JAVSO..40..539C,2013A&A...551A..77T}.
Two scenarios have been proposed for the origin of an RCB star: the double
degenerate (DD) and the final helium-shell flash (FF) models \citep{Iben:1996fj}. The former involves the
merger of a CO- and a He-white dwarf (WD) \citep{1984ApJ...277..355W}.
 In the latter, a star evolving into a planetary nebula
central star is blown up to supergiant size by a FF \citep{1979sss..meet..155R}.

Their defining characteristic is unusual variability - RCB stars
undergo massive declines of up to 8 mag due to the formation of carbon
dust at irregular intervals. 
\citep{1996PASP..108..225C}. 
Typically the atmospheres of these stars are, $\sim$98\% helium, $\sim$1\% carbon, and almost no 
hydrogen \citep{Asplund:2000qy}. It has been suggested that when 
dust forms around an RCB star, radiation pressure accelerates the dust 
away from the star, dragging gas along with it 
\citep{1963ApJ...138..320P,1992ApJ...397..652C}.  There is evidence for 
both low (10-20 km s$^{-1}$) and high (200-400 km s$^{-1}$) velocity gas 
streaming away from some RCB stars 
\citep{1992ApJ...397..652C,2003ApJ...595..412C,2011ApJ...739...37A}.
However, it is not known if wind speeds and mass loss rates are 
time-variable and if so how they depend on dust formation events.
 
\citet{2003ApJ...595..412C} presented velocity-resolved spectra of the 
He~I 2$^{3}P-2^{3}$S $\lambda$10830 line in RCB stars showing that most, 
if not all of them have winds. Nine of the ten stars observed 
showed P-Cygni or asymmetric blue-shifted line profiles. As ionization 
of He~I requires photons of energies $>$ 24~eV it is likely that the $n$ 
= 2 levels of He~I that produce this line are populated by collisions 
rather than by photoionization/recombination in these stars, whose 
effective temperatures imply very low far ultraviolet luminosities. The 
observed $\lambda$10830 line profiles were modeled using a Sobolev plus 
Exact Integration (SEI) code to derive optical depths and velocity 
fields of the helium \citep{2003ApJ...595..412C}. The fits imply that 
the helium in a typical RCB wind undergoes steep acceleration to a 
terminal velocity of $\Vinf$ = 200-350 \kms and has a column density of 
N$_{He}$ $\sim$10$^{12}$ cm$^{-2}$ in the metastable 2$^{3}$S level. 
Interestingly, the five known hydrogen-deficient carbon (HdC) stars, 
which are spectroscopically identical to the RCB stars, but produce no 
dust (i.e., do not vary in optical brightness), have also been
observed at He I $\lambda$10830 \citep{2009ApJ...698..735G}. No evidence 
for a stellar wind was found in any of the five stars, consistent with 
the suspected dust -- wind connection in this extended family of stars.

\citet{2003ApJ...595..412C} suggested that there may be a
relationship between the light curve of an RCB star and its He I 
$\lambda$10830 profile, as the He I wind features were stronger when they were
closer in time to a dust formation episode. Here we report 
on five epochs of He I $\lambda$10830 spectra of twelve RCB stars, 
obtained over a three year period. We use these data to study how the 
equivalent widths and terminal velocities of the line vary with time and to 
determine more quantitatively if those variations are related to the 
timing of dust formation episodes.
 
\section{Observations and Data Reduction}

Observations of He I $\lambda$10830 in twelve RCB stars were obtained on 
six nights of highly variable weather conditions between 2001 June and 
2004 July at the United Kingdom Infrared Telescope (UKIRT) using its 
facility spectrometer CGS4 configured with its echelle grating and 
two-pixel (0\farcs9) wide slit. The observations are summarized in Table 
1. The resolution was 5 $\times 10^{-5}$ \micron\ (14 km~s$^{-1}$). The 
spectra were ratioed with those of nearby A stars in order to remove 
telluric features. Wavelength calibration was achieved using telluric 
absorption lines observed in the comparison stars. A quadratic fit was 
made to a selection of these lines covering the entire spectral range. 
The 1-$\sigma$ wavelength uncertainty is 5-8 $\times 10^{-6}$ \micron. 
The spectra have been corrected for each star's radial velocity, and a 
heliocentric correction has also been made based on the epoch of each 
observation. Thus, in the final reduced spectra the wavelength scale 
({\it in vacuo}) refers to the rest frame of the star and the wavelength 
of the He I $\lambda$10830 line is 1.08333~$\mu$m 
\citep{2009ApJ...698..735G}. One star, ES Aql, does not have a published 
radial velocity. A value of -70 km~s$^{-1}$ has been inferred for it 
from the spectra analyzed here. The stellar radial velocities used for this 
study are listed in Table 1.

The normalized spectra, slightly smoothed to a resolution of 6.5 $\times 
10^{-5}$ \micron~(18 km~s$^{-1}$), are shown in Fig.~1. 
In most cases the helium line dominates the spectral region. The wavelengths of the other prominent atomic lines are indicated in the figure.  Many other absorption lines are present. The signal-to-noise ratio exceeds 50 in most cases, but in some spectra (V854 Cen, 3rd spectrum from top, (UX Ant, all spectra; WX CrA, top spectrum; V517 Oph, top spectrum; U Aqr, all spectra; ES Aql, top spectrum; SV Sge, top spectrum) the noise level can be estimated from the fluctuations between strong spectral features.

As listed in 
Table 1, the RCB stars range in effective temperature (T$_{eff}$) from 
4000 to 7500~K. Their spectra have very different appearances depending 
on whether T$_{eff}$ is greater than or less than $\sim$6000 K 
\citep{Asplund:2000qy}. The warmer stars, plotted in Figure 1a, show 
mainly atomic absorptions of C I and singly ionized metals while the 
cooler stars, in Figure 1b, show in addition, strong C$_2$ and CN 
absorption bands. The line identifications are from 
\citet{1974PASJ...26..163H}, and \citet{1995PASP..107.1042H}. Two stars, 
ES Aql and V517 Oph, do not have reported values of T$_{eff}$, but their 
spectra resemble those of RCB stars with T$_{eff}$ $\sim$ 5000 K. Note 
that each panel in Fig. 1 contains a model RCB stellar spectrum, of 
either a 5400 K or a 6500 K RCB star \citep{2009ApJ...698..735G,2009ApJ...696.1733G} depending on which is 
more appropriate to the effective temperature of the star.

Light curves of the sample stars for the time period of the He~I 
observations are shown in Fig.~2. They consist of visual data downloaded 
from the AAVSO International Database (Henden, A.A., 2008, private 
communication), and V-band data from the ASAS-3 telescope 
\citep{2002AcA....52..397P}. The epochs at which He I $\lambda$10830 
spectra were obtained for each star, listed in Table 1, are marked on 
the light curves.

The equivalent widths (EW)and terminal velocity (v$_\infty$) of the blue-shifted absorption in the He I 
$\lambda$10830 line were measured for each spectrum. They are listed in Table 2. The wavelength 
range for which the EW was calculated was determined by comparing each 
spectrum to model RCB stellar spectra. The EW was corrected to the 
extent possible for stellar absorption lines, in particular for the Si I 
line at 1.0830 \micron. In a few spectra, an absorption could not be 
measured because strong emission extended over a wide velocity range 
and/or the continuum was too weak to reliably measure absorption. For 
all other spectra, the absorption EW and v$_\infty$ for each epoch is plotted against the time in days 
since the end of that star's latest decline (also listed in Table 2) in Fig.~3, with the values 
of those quantities during the decline shown on the left hand edges of 
the plots. Note that the filled symbols in Fig. 3 comprise the data for 
all but two of the stars, whereas the open symbols apply to two unusual 
cases, V482 Cyg and V854 Cen.

\section{Discussion}

\subsection{Previous Observations of He I $\lambda$10830}

Blue-shifted high-velocity absorption features (100--400 km~s$^{-1}$) 
have been reported from time to time in the spectra of RCB stars, 
observed at early and late times during declines 
\citep{1972MNRAS.158..305A,1990MNRAS.244..149C,1992ApJ...384L..19C,1993MNRAS.264L..13C,1994ApJ...432..785C,1995PASP..107..244V,1997PASP..109..796G,Rao:1997}. 
For example, the Na I D lines in R CrB showed blue-shifted absorption 
during that star's recovery from its 1995-96 decline \citep{Rao:1999}. 
The line profiles implied that the gas was being accelerated from 100 to 
200 km~s$^{-1}$. During its 2003 minimum, R CrB showed blue wings in the 
infrared O I triplet lines implying ejection at 130 km~s$^{-1}$ 
\citep{Kameswara-Rao:2006lr}. However, a blue wing was also present at 
maximum light. A similar ejection velocity in V CrA was also determined 
from its O I line profiles \citep{Rao:2008lr}.

Prior to the small survey by \citet{2003ApJ...595..412C}, very few 
observations of the He I $\lambda$10830 line in RCB stars had been 
reported. R CrB, itself, however, was observed at five epochs between 
1972 and 1978, and showed a P-Cygni or asymmetric blue-shifted profile 
at all times, whether in decline or at maximum light 
\citep{Wing:1972qy,Querci:1978fk,Zirin:1982uq}. R CrB 
showed emission at He I $\lambda$10830 in three spectra obtained during 
a decline in 1972 March--May when it was between V = 9 and 12 mag 
\citep{Wing:1972qy}. On 1977 January 18 at V =7.0, during recovery after 
a decline, the line had a P-Cygni profile with v$_{\infty}$= -240 
km~s$^{-1}$ \citep{Querci:1978fk}. On 1978 July 24, R CrB showed very 
strong absorption at He I $\lambda$10830 with a velocity shift of 
$\sim$-200 km~s$^{-1}$, several months after it had returned to maximum 
light after a decline \citep{Zirin:1982uq}. The helium-final-flash star, 
Sakurai's Object, which may be related to the RCB stars, showed an 
increasingly blue-shifted absorption out to 670 \kms~at He I 
$\lambda$10830 as the star formed dust and went into a deep RCB-like 
decline \citep{Eyres:1999qy}, from which it has yet to recover.

\subsection{New Observations}

Temporal correlations between dust obscuration and He I $\lambda$10830 line 
properties associated with wind activity can be discerned by comparing 
the individual observed spectra in Fig. 1 to the individual light curves 
in Fig. 2, or collectively by examination of Fig. 3, in which the key 
line parameters, absorption strength, and terminal velocity are 
presented for all of the stars on a single time axis whose origin is the 
end of the most recent decline of each star. Viewed collectively, as in 
Fig. 3, the data reveal a fairly uniform behavioral pattern. There is a 
strong correlation between visual obscuration and the strength of the 
blue-shifted He I $\lambda$10830 absorption, with increased obscuration 
associated with a large increase in the line strength. There also is a 
clear tendency for the maximum velocity of the wind to be greater during 
a decline than well after it has ended.

Even small episodes of dust formation, such as that seen in V CrA on 
$\sim$JD 2452800 (Fig.~2), are coincident with significant increases in 
wind activity. V CrA had declined visually by only $\sim$0.5 mag but the 
EW of the wind features (the 2nd and 3rd spectra from the top in Fig. 1) 
increased greatly. There is some indication that the cooler RCB stars in 
the sample, V517 Oph, ES Aql, SV Sge and WX CrA have larger equivalent 
widths in He I during declines than the warmer stars. However, this 
might be due to sampling statistics, as a higher fraction of He I 
spectra of cool RCB stars were obtained during deep declines than for 
the warm RCB stars. There is also a tendency for a star in a deep 
decline, such that the stellar continuum is weak relative to the He I 
line emission, to show a classic P-Cygni profile in the $\lambda$10830 
line. This can be seen in Fig. 1b for WX CrA and U Aqr. In the last two spectra of SV Sge, the star was deep in a decline, and so had no continuum for an absorption to be seen against, but it showed strong He I $\lambda$10830 emission.

Figure 3 reveals that the strength of the wind absorption drops off 
sharply beginning about 100 days after the end of a decline; that is, 
well after the star has returned to maximum brightness and there is 
ostensibly no dust along the line of sight. The same behavior is seen in the maximum 
velocity of the wind; it does not decrease significantly from its value 
during the decline of 300--400~km~s$^{-1}$ until after $\sim$100 days. The 
latter behavior may not be surprising as high velocity gas far from the 
star may be expected to decelerate slowly. However, as most of the 
absorption is expected to occur in the densest part of the wind, closest 
to the star, one might predict the drop-off in absorption strength to 
commence very quickly after dust formation ceases and to decrease 
gradually thereafter.

Note, however, that even small episodes of dust formation profoundly 
affect the wind properties, as seen in V CrA. Thus it may be that 
sufficient dust continues to form and accelerate the gas during this 
$\sim$100-day interval. 
It is sometimes difficult to distinguish between pulsations and small dust formation episodes \citep[e.g.,][]{1993MNRAS.265..899F,1995PASP..107..416C}.
More accurate photometry might be able to reveal 
this.

An additional noteworthy result is that a weak blueshifted He I $\lambda$ 
10830 line is present and roughly constant in strength in all of the 
stars in the sample even well beyond 100 days after each stars most 
recent decline.  This implies the existence of a weak high velocity wind 
at all times and generalizes the conclusion drawn from historical 
observations of R CrB that a wind is always present, as discussed 
earlier. The constancy of this residual wind may suggest that a small amount 
of dust formation is always taking place.

The He~I spectra in our sample that were obtained, by chance, just before a 
decline are sparse, but suggest that there are no precursive indications 
of changes in the line prior to a decline. The best examples are the 
two spectra of RY Sgr at $\sim$JD 2453150 (about 100 days before a 
decline), neither of which shows an increasing wind feature.

There are two exceptions to the general behavior described 
above: V854 Cen and V482 Cyg, whose line parameters are plotted 
separately from the other RCB stars in Fig. 3. V854 Cen, in particular, 
appears to behave differently. It had more dust formation episodes than 
any of the other RCB stars during 2001--2004. One might have expected, 
based on the behavior of the majority stars, that its wind features 
would therefore be very strong. Instead the opposite was observed: its 
spectrum has almost no He I $\lambda$10830 absorption at the end of one 
decline ($\sim$JD 2452100), has modest absorption (albeit highly 
blueshifted) at the end of another decline ($\sim$JD 2452800), and again 
little or no absorption a month later. Unfortunately, no line 
measurements of V854 Cen were made when deep in a decline. In Figure 3, it can be 
seen that V854 Cen's terminal velocity, when measurable, is considerably 
higher than the other RCB stars. The rapid disappearance of its He I
$\lambda$10830 absorption after a dust formation episode could be 
attributed in part to that. V854 Cen is often the odd man out among RCB 
stars \citep{1996PASP..108..225C,2012JAVSO..40..539C}. It has a relatively large hydrogen 
abundance even though it is still hydrogen deficient, and has unusual 
abundances for an RCB star \citep{1998A&A...332..651A, Asplund:2000qy}.

V482 Cyg is the only star in the sample to show no dust formation activity 
during the epoch of He I observations, and it was already 1100 d past its last 
decline when the first He I $\lambda$10830 spectra were obtained. Its 
line has the smallest measured EW in the sample. 
V482 Cyg is
an extreme example of the general behavior, 
showing that a small residual wind is present even years after the latest decline.

\subsection{Dust-Driven Winds in RCB Stars}

In this section, we discuss the origin of the winds in RCB stars. We use simple equations to derive estimates of the mass-loss rate and wind terminal velocity, and compare those estimates to our own and other results, showing that models in which dust accelerates the gas and collisionally excites the lowest levels of the helium triplet state are consistent with the observations.

The lower state of the He I $\lambda$10830  (2$^3$S - 2$^3$P) transition is more than 20 eV above 
the ground state. The transition probability is
very small (A$_{21}$=1.27 x 10$^{-4}$ s$^{-1}$) \citep{1994ApJ...423..785S}.  
This metastable state could be populated
by one of two mechanisms, photoionization/recombination or collisional excitation.
He II $\lambda$1640 is not seen in RCB stars which indicates that the He I $\lambda$10830 line is not being formed by
helium photoionization and recombination  \citep{1994AJ....107.1128R,1996PASP..108..225C,1999AJ....117.3007L}.
R CrB, itself, showed emission lines of
 He I $\lambda\lambda$ 3889, 5876, 7065 and 10830 \AA\
during the 1995-96 decline of R CrB \citep{Rao:1999}.
At the time of the observations, the 5876 \AA\ line was much weaker than the 3889 and 7065 \AA\ lines, which makes sense if
the lines are optically thick and the electron 
density is high.
\citet{Rao:1999} estimate T $\sim$ 20000 K and $n_e$ = 10$^{11}$--10$^{12}$ 
cm$^{-3}$ so collisional excitation is important.
In the final-flash star, Sakurai's object, thought to be similar to RCB stars, the He I 
$\lambda$10830 emission may be
the result of collisional excitation in shocked gas being dragged outward by 
the expanding 
dust cloud
around the star \citep{Eyres:1999qy,2000MNRAS.315..595T}.
The kinetic energy of a 400 km s$^{-1}$ He atom is $\sim$3200 eV, which is 160 times the excitation energy of the 2$^3$S and 2$^3$P triplet levels, so as the He is accelerated by the dust, 20 eV collisional energy transfers exciting it to the triplet states are quite plausible. Collisions with relative velocities $>\sim$30 km s$^{-1}$ will have enough energy to excite the He atoms.

Dust can form anywhere around an RCB star, but a decline in brightness 
only occurs when dust forms along the line of sight. The blue-shifted 
absorption seen in He I $\lambda$10830 is created in gas that also is 
along the line of sight. The two phenomena are thus caused by gas and dust located together in 
the same column toward the star and it is quite likely that they are related. An inspection of 
Fig. 2 shows that the RCB stars in the sample had a wide range of dust 
formation activity during the time period of the He I measurements. Some 
RCB stars are more active in general than others, but each star also 
shows changes in its level of dust formation activity on timescales no 
longer than several years 
\citep{1996AcA....46..325J,1996PASP..108..225C}.

Generally, the maximum possible mass-loss rate, $\dot{M}_o$,  due to radiation pressure on dust grains is,

\begin{equation}
\dot{M}_o=\frac{L_\star}{cv_{\infty}}
\end{equation}

where L$_\star$ is the stellar luminosity and v$_{\infty}$ is the terminal velocity \citep[e.g.,][]{1982ApJ...252..616K}. But, if the optical depth of dust is large, then multiple scattering becomes important and,

\begin{equation}
\dot{M}=\dot{M}_o \tau_{\infty}
\end{equation}

where  $\tau_{\infty}$ is the total optical depth of the circumstellar dust \citep{1986A&A...161..201G}.

A warm RCB star like R CrB, itself, with M$_V$ = --5 mag, has L$_\star$ $\sim$10$^4$ L$_\sun$ and R$_\star$ $\sim$ 70 R$_\sun$ \citep{2012JAVSO..40..539C}. From Figure 3, a typical terminal velocity is v$_{\infty}$= 400 km s$^{-1}$. So $\dot{M}_o$ $\sim$ 5 $\times 10^{-7}$ M$_{\sun}$ yr$^{-1}$. An RCB star decline can be 8 mag below maximum at V, so assuming $\tau_{\infty}$ $\sim$ 8, then $\dot{M}$=4 $\times 10^{-6}$ M$_{\sun}$  yr$^{-1}$. This agrees well with estimates of RCB star mass loss from observations of R CrB, itself \citep[][and references therein]{2011ApJ...743...44C}.

The radiation pressure on a spherical dust grain is,
\begin{equation}
F_{pr}=\frac{L_\star a^2Q_{pr}}{4cr^2}
\end{equation}

where Q$_{pr}$ is the radiation pressure efficiency and r is the distance from the center of the star  \citep{1976ApJ...203..552M}. Then, integrating from r = r$_o$, the condensation radius of the dust,  to r=$\infty$ and from v= 0 to v=v$_{\infty}$ produces,

\begin{equation}
v_{\infty}=\left(\frac{3\left(\frac{m_d}{m_g}\right)L_\star Q_{pr}}{8\pi a \rho_s c r_o}\right)^{1/2}
\end{equation}

where $\frac{m_d}{m_g}$ is the mass dust-to-gas ratio, and a and $\rho_s$ are the radius and density of the dust grains. The inclusion of the dust-to-gas ratio takes into account the momentum transfer from the dust to the much more massive He gas. RCB stars are $\sim$98\% He \citep{Asplund:2000qy}.

If the dust is amorphous carbon, then $\rho_s$ $\sim$ 2 g cm$^{-3}$. 
Assuming that Q$_{pr}$ = 1, and r$_o$ = 2 R$_\star$ = 1 $\times$ 10$^{13}$ cm \citep{1996PASP..108..225C}, and a typical dust-to-gas ratio for the Galaxy of 0.01 \citep{2007ApJ...663..866D}, then,
v$_{\infty}$= 2.8 a$^{-1/2}$. So, if a dust grain radius is $\sim$0.5 \micron, then v$_{\infty}$ $\sim$ 400 km s$^{-1}$, which is typical for v$_\infty$ measured from  He I $\lambda$10830 in this paper. This is slightly larger than other estimates of grain sizes around RCB stars \citep[e.g.,][]{1985MNRAS.217..767E,1988MNRAS.233...65F}. But the dust grain size inferred here is consistent with other estimates considering that other parameters in Equation 4, such as the dust-to-gas ratio, the stellar luminosity, and the dust condensation radius are also somewhat uncertain.

\section{Conclusions}

The current data, although difficult to interpret for individual RCB 
stars due to the small number of observations per star, collectively they
show a consistent behavioral pattern for the overwhelming majority of 
stars in the sample. At maximum light, most RCB stars show weak and highly 
blue-shifted ($\sim$200 km s$^{-1}$) He I $\lambda$10830 line absorption 
from the excited 2$^3$S state, while at the same time producing a
photospheric spectrum that is unshifted in the rest-frame of the star.  Each decline in optical brightness 
coincides with an large increase in the He I $\lambda$10830 line EW,
coupled with an increase in v$_\infty$ of the wind to 
$\sim$400 km~s$^{-1}$. Following recovery to the pre-decline brightness, 
both the higher v$_\infty$ and the increased EW
persist for $\sim$3 months before returning to their original pre-decline values.
This strong correlation between dust formation and the blue-shifted absorption in He I $\lambda$10830 is supported by the observations of the HdC stars, which are identical to the RCB stars spectroscopically, but produce no dust and have no evidence for blue-shifted absorption in He I $\lambda$10830 \citep{2009ApJ...698..735G}. So, in the RCB stars, helium gas is dragged along by dust grains accelerated by radiation pressure from the star. 

The close connection between dust formation and this type of wind 
behavior provides strong support for the long held view that radiation 
pressure accelerates the newly formed dust and that it in turn 
collisionally excites the gas.  The escape velocity from RCB is 
estimated to be 30-70 km~s$^{-1}$ \citep{Rao:1997}. Thus, the observed 
maximum velocities greatly exceed the escape velocities from 
these stars. A simple explanation for the presence of a (much weaker) 
wind in between declines is that a lower level of dust formation than 
observed in the declines is always taking place.

\citet{2011ApJ...739...37A} have 
found evidence in the optical spectrum of scattered light from R CrB for 
dust velocities of only $\sim$25 km~s$^{-1}$. That dust cannot be 
connected with the much higher velocities inferred from the 
He I $\lambda$10830 line profile. The scattering may occur in dust located 
much further from the star than the dust responsible for the 
simultaneous strong obscuration of the starlight and acceleration of the 
gas, perhaps where the dust either has been greatly decelerated or where 
large quantities of dust ejected long ago at lower velocities have 
accumulated.  R CrB has a large dust shell around it that may be a fossil planetary nebula shell 
\citep{2011ApJ...743...44C}.
In view of the above estimate of escape velocity, gas and 
dust ejected at low velocities from the star probably would not travel 
far before being significantly decelerated and perhaps fall back into 
the star.

Additional observations of the He $\lambda$10830 line, probably in the 
form of a more intensive campaign than the one reported here, combined 
with more accurate photometry are now needed to more stringently test 
and refine our conclusions.

\acknowledgments

We thank the anonymous referee for several helpful comments. We thank the staff of the United Kingdom Infrared Telescope, which is 
operated by the Joint Astronomy Centre on behalf of the U.K. Particle 
Physics and Astronomy Research Council. We acknowledge with thanks the 
variable star observations from the AAVSO International Database 
contributed by observers worldwide and used in this research. TRG's 
research is supported by the Gemini Observatory, which is operated by 
the Association of Universities for Research in Astronomy, Inc., on 
behalf of the international Gemini partnership of Argentina, Australia, 
Brazil, Canada, Chile, the United Kingdom, and the United States of 
America. 


\bibliography{/Users/gclayton/projects/latexstuff/everything2}

\begin{deluxetable}{lrrcccccc}
\tablecolumns{7}
\tablewidth{0pc}
\tablecaption{UKIRT Observations}
\tablenum{1}
\tablehead{\colhead{Star} & \colhead{T$_{eff}^a$} & \colhead{RV$^b$} &\colhead{2001} & \colhead{2003}& \colhead{2003}& \colhead{2004}& \colhead{2004}& \colhead{2004}\\
\colhead{} &\colhead{} &\colhead{} &\colhead{June 15} &\colhead{May 2} &\colhead{June 2} &\colhead{May 6} &\colhead{June 4} &\colhead{July 17}
}

\startdata
RY Sgr&7250&-20.5&X&X&X&X&X&\\
UX Ant&7000&+144.3&&&X&X&X&\\
R CrB&6750&+22.3&X&X&X&X&X&\\
V2552 Oph&6750&+60.5&&&X&X&X&\\
V854 Cen&6750&-25.1&X&X&X&X&X&\\
V482 Cyg&6500&-41.2&X&X&X&X&X&\\
V CrA&6250&-7.9&X&X&X&X&X&\\
U Aqr&6000&+98.0&X&&X&X&X&X\\
WX CrA&5300&-3.8&X&&&X&X&\\
ES Aql&$\sim$5000&-70.0&X&X&X&X&X&\\
V517 Oph&$\sim$5000&+21.0&X&X&X&X&X&\\
SV Sge&4000&+4.0&X&X&X&X&X&\\
\enddata
\tablenotetext{a}{\citep{1997A&A...318..521A,1998A&A...332..651A,Asplund:2000qy,2001A&A...369..178B,Rao:2003vn}. The effective temperatures of ES Aql and V517 Oph have never been calculated, but both show spectra indicative of cool RCB stars.}
\tablenotetext{b}{\citep{1953GCRV..C......0W,Lawson:1989kx,Kilkenny:1992fk,Rao:1993yq,Lawson:1997fj,Rao:2003vn}}
\end{deluxetable}

\begin{deluxetable}{lllll}
\tablecolumns{5}
\tablewidth{0pc}
\tablecaption{Measurements}
\tablenum{2}
\tablehead{\colhead{Star} & \colhead{Epoch$^a$} & \colhead{Days$^b$} &\colhead{EW (\AA)} & \colhead{V$_\infty$}}
\startdata
RY Sgr	&	1	&	0	&	5.5	&	-345	\\
	&	2	&	686	&	1.2	&	-258	\\
	&	3	&	717	&	1.0	&	-259	\\
	&	4	&	1056	&	1.0	&	-250	\\
	&	5	&	1085	&	1.2	&	-257	\\
UX Ant	&	3	&	0	&	4.5	&	-420	\\
	&	4	&	231	&	2.3	&	-408	\\
	&	5	&	260	&	2.6	&	-417	\\
R CrB	&	1	&	126	&	5.1	&	-392	\\
	&	2	&	11	&	6.6	&	-427	\\
	&	3	&	42	&	4.9	&	-413	\\
	&	4	&	381	&	1.1	&	-313	\\
	&	5	&	410	&	0.9	&	-303	\\
V2552 Oph	&	3	&	292	&	1.8	&	-182	\\
&	4	&	631	&	2.1	&	-198	\\
	&	5	&	660	&	1.9	&	-186	\\
V854 Cen	&	1	&	0	&	1.0	&	-716	\\
	&	2	&	11	&	5.1	&	-605	\\
	&	3	&	42	&	2.0	&	-611	\\
	&	4	&	381	&	1.3	&	-718	\\
	&	5	&	410	&	0.7	&	-604	\\
V482 Cyg	&	1	&	1376	&	0.9	&	-225	\\
	&	2	&	1476	&	0.0	&	\nodata	\\
	&	3	&	1507	&	1.0	&	-179	\\
	&	4	&	2431	&	0.9	&	-179	\\
	&	5	&	2460	&	0.9	&	-218	\\
V CrA	&	1	&	826	&	1.3	&	-366	\\
	&	2	&	0	&	6.0	&	-397	\\
	&	3	&	0	&	6.4	&	-435	\\
	&	4	&	131	&	1.5	&	-346	\\
	&	5	&	160	&	1.1	&	-304	\\
U Aqr	&	1	&	0	&	4.3	&	-238	\\
	&	3	&	0	&	6.3	&	-285	\\
	&	4	&	0	&	4.6	&	-288	\\
	&	5	&	0	&	5.4	&	-275	\\
	&	6	&	0	&	5.8	&	-276	\\
WX CrA	&	1	&	0	&	7.7	&	-308	\\
	&	4	&	0	&	5.6	&	-382	\\
	&	5	&	0	&	6.6	&	-383	\\
ES Aql	&	1	&	0	&	5.2	&	-367	\\
	&	2	&	0	&	10.2	&	-444	\\
	&	3	&	0	&	10.2	&	-432	\\
	&	4	&	0	&	11.0	&	-400	\\
	&	5	&	0	&	11.2	&	-433	\\
V517 Oph	&	1	&	0	&	9.8	&	-507	\\
	&	2	&	61	&	6.2	&	-409	\\
	&	3	&	92	&	6.9	&	-412	\\
	&	4	&	0	&	9.3	&	-414	\\
	&	5	&	0	&	10.7	&	-510	\\
SV Sge	&	1	&	0	&	5.8	&	-312	\\
	&	2	&	161	&	3.4	&	-315	\\
	&	3	&	192	&	5.3	&	-313	\\
	&	4	&	0	&	\nodata	&	\nodata	\\
	&	5	&	0	&	\nodata	&	\nodata	\\
\enddata
\tablenotetext{a}{The epoch numbers refer to the six epochs in Table 1.}
\tablenotetext{b}{Days since the end of the last decline.}
\end{deluxetable}

\begin{figure} \figurenum{1a} 
\includegraphics[width=3.25in,angle=0]{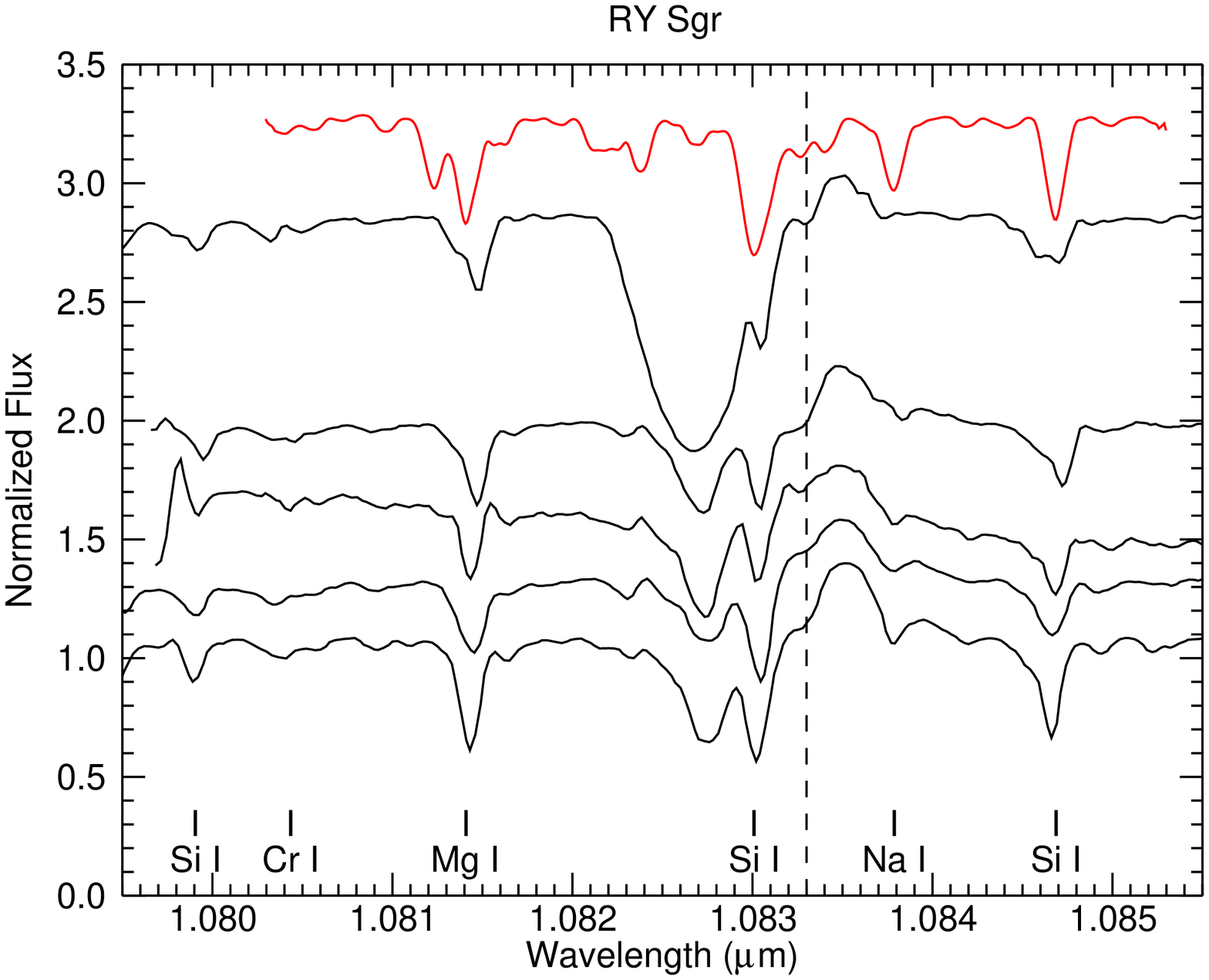} \hspace{0.5in} 
\includegraphics[width=3.25in]{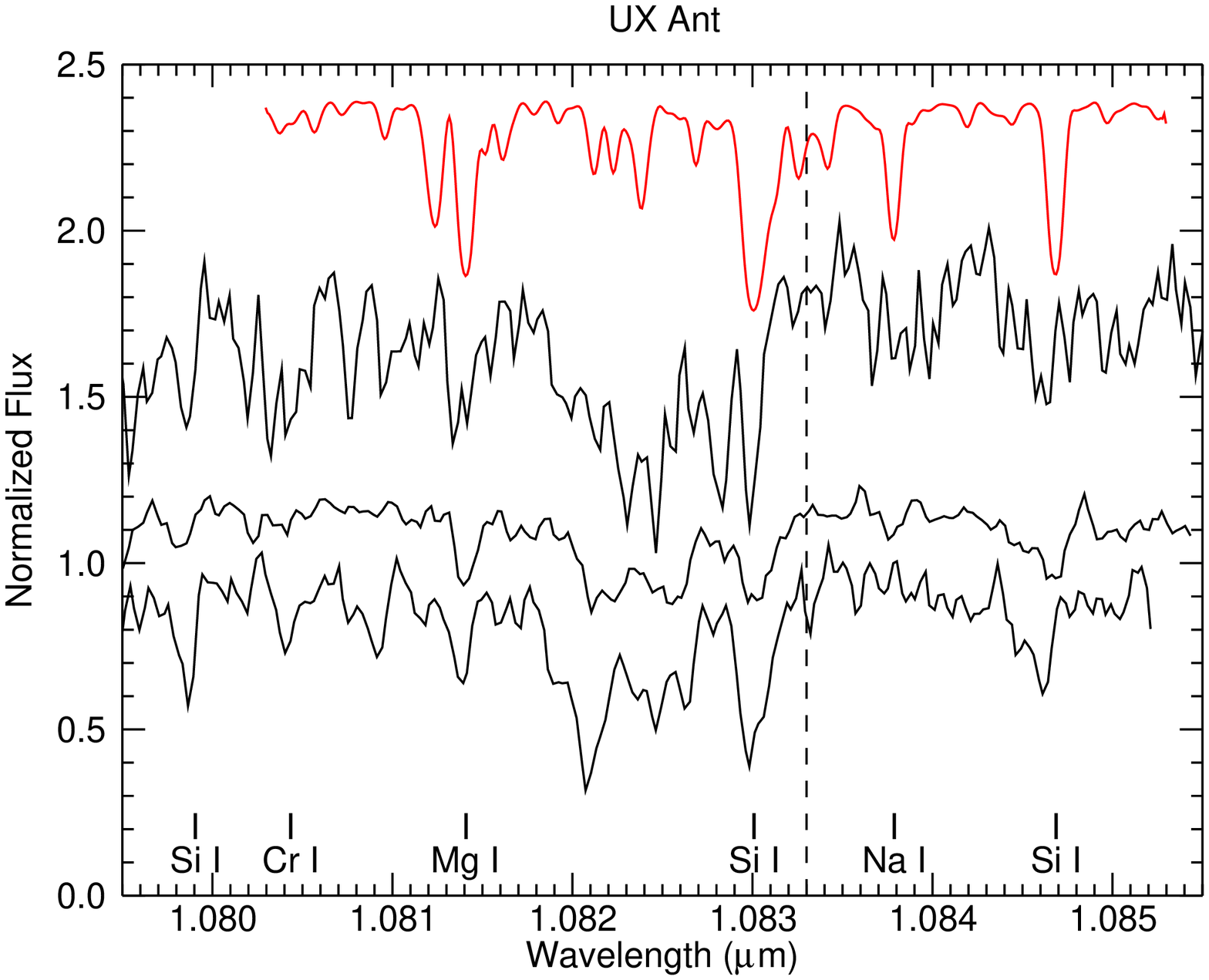} 
\includegraphics[width=3.25in,angle=0]{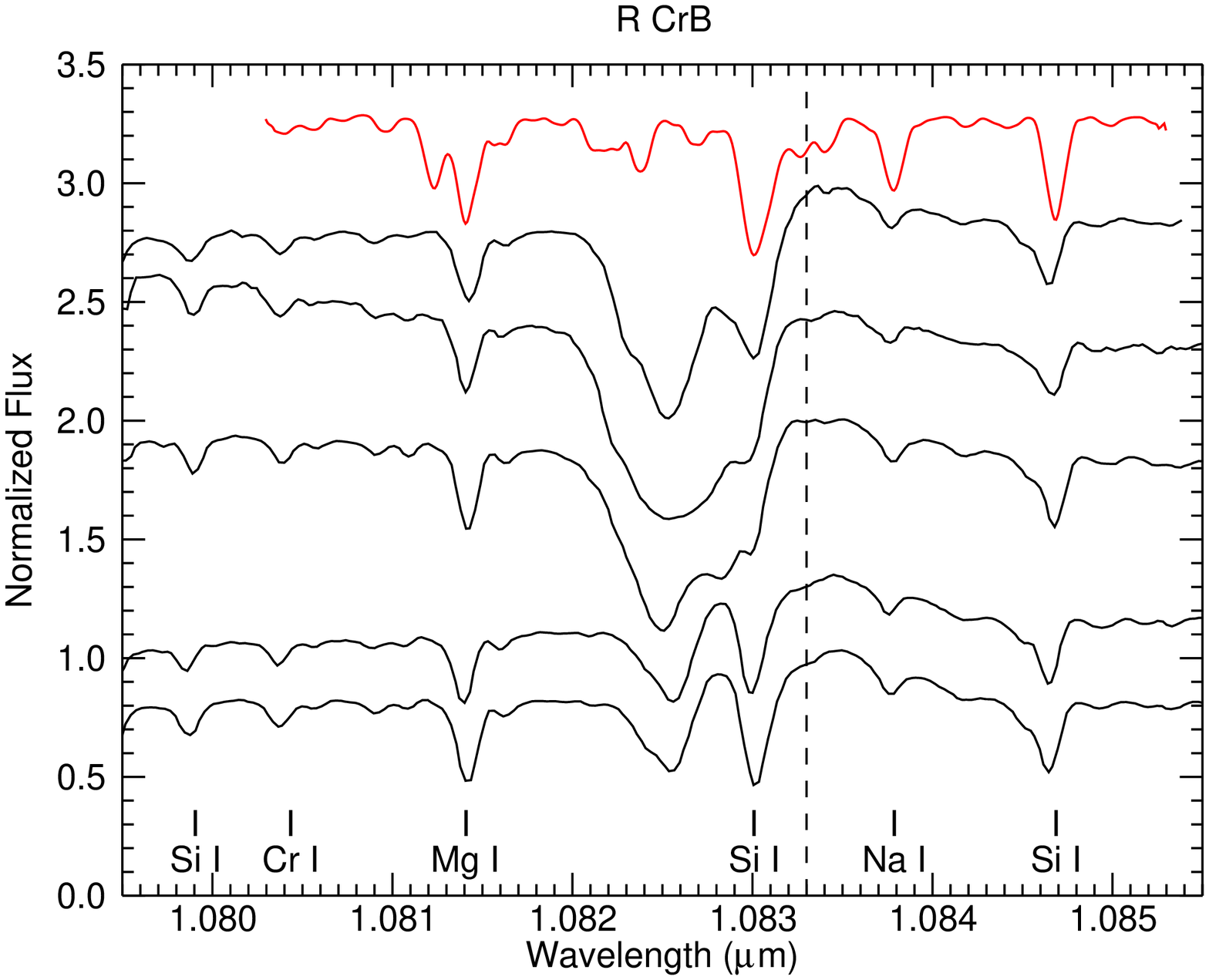} \hspace{0.5in} 
\includegraphics[width=3.25in]{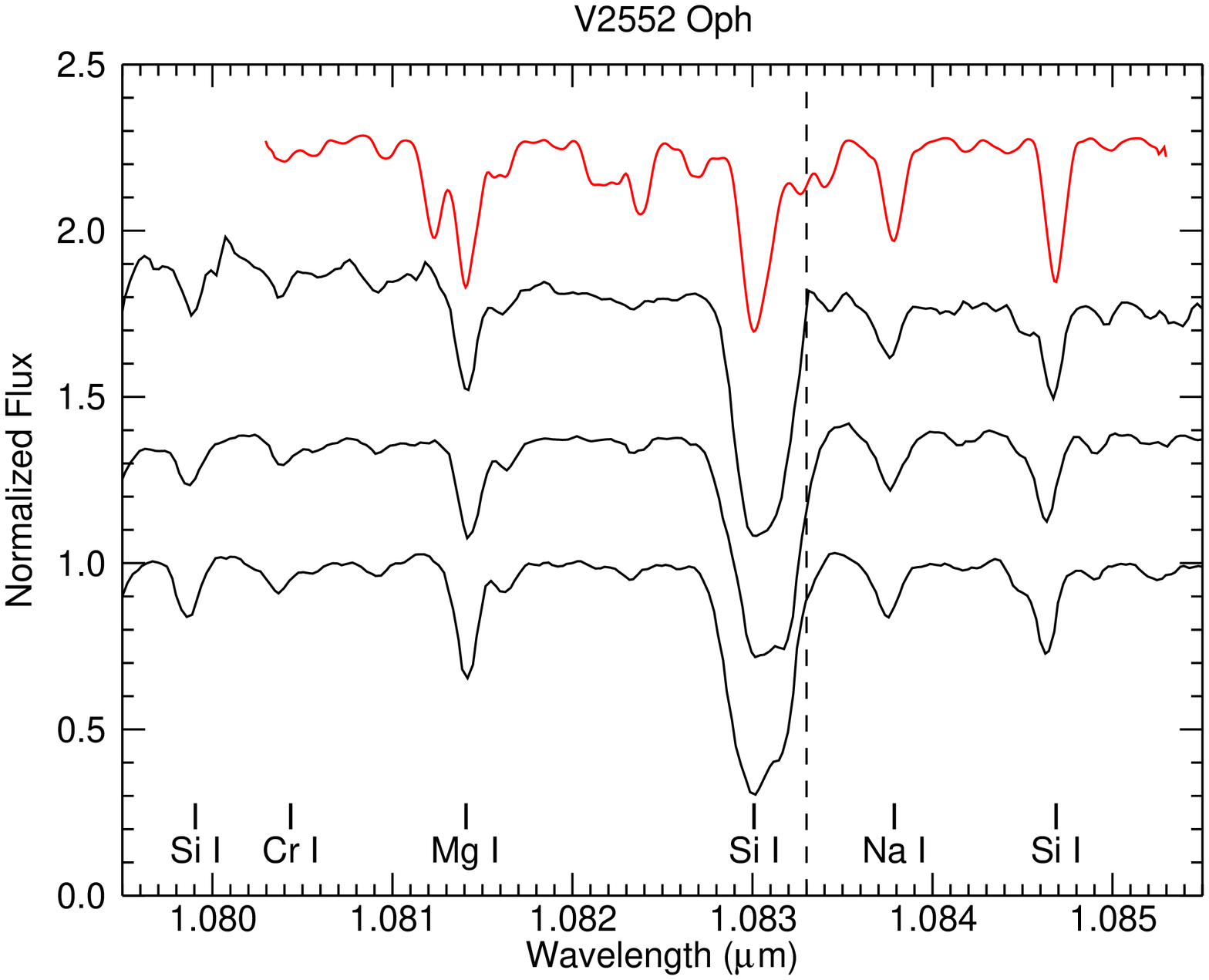} 
\includegraphics[width=3.25in]{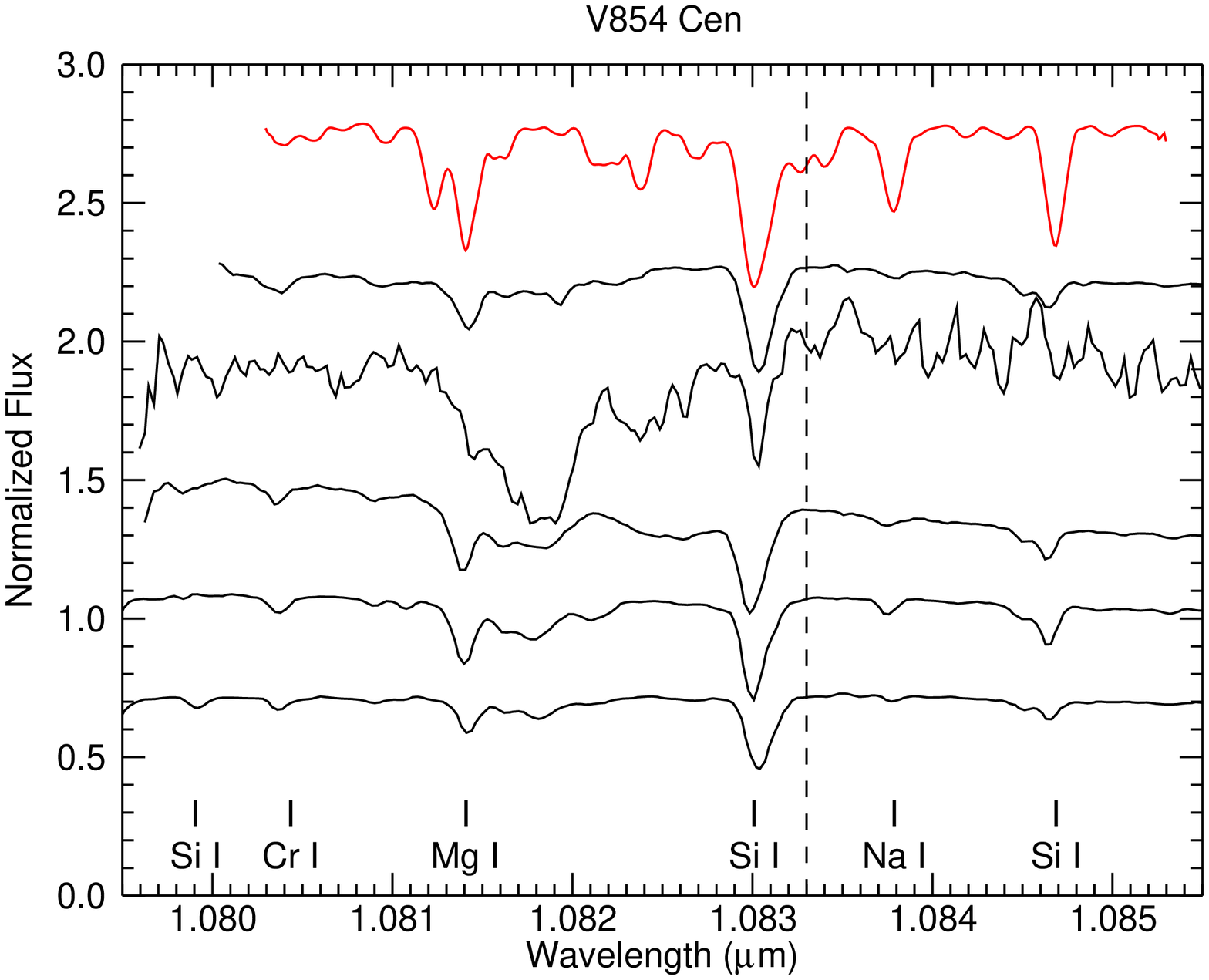} \hspace{0.5in} 
\includegraphics[width=3.25in]{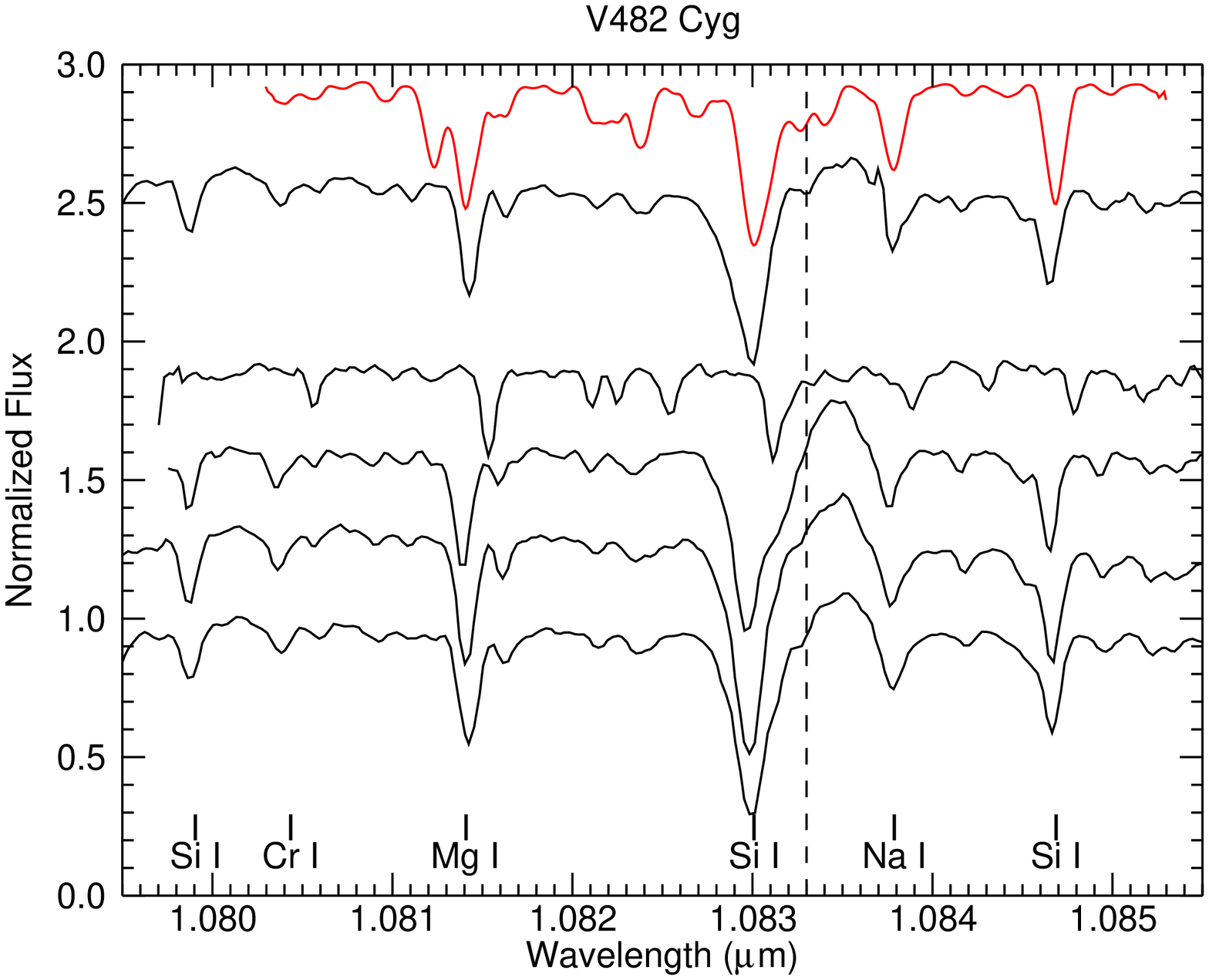} 
\caption{UKIRT CGS4 echelle spectra in the region of the He I 
$\lambda$10830 line. Vacuum wavelengths are plotted. 
The vertical dashed line shows the wavelength, 1.08333~$\mu$m,
of the He I $\lambda$10830
\citep{2009ApJ...698..735G}.
In each panel 
individual normalized spectra are shifted vertically and are in 
chronological order from top to bottom. The epochs for each star are 
listed in Table 1. The red spectrum at the top of each panel is a 6500 K 
RCB stellar model \citep{2009ApJ...698..735G,2009ApJ...696.1733G}.} 
\end{figure}

\begin{figure}
\figurenum{1b} 
\includegraphics*[width=3.25in,angle=0]{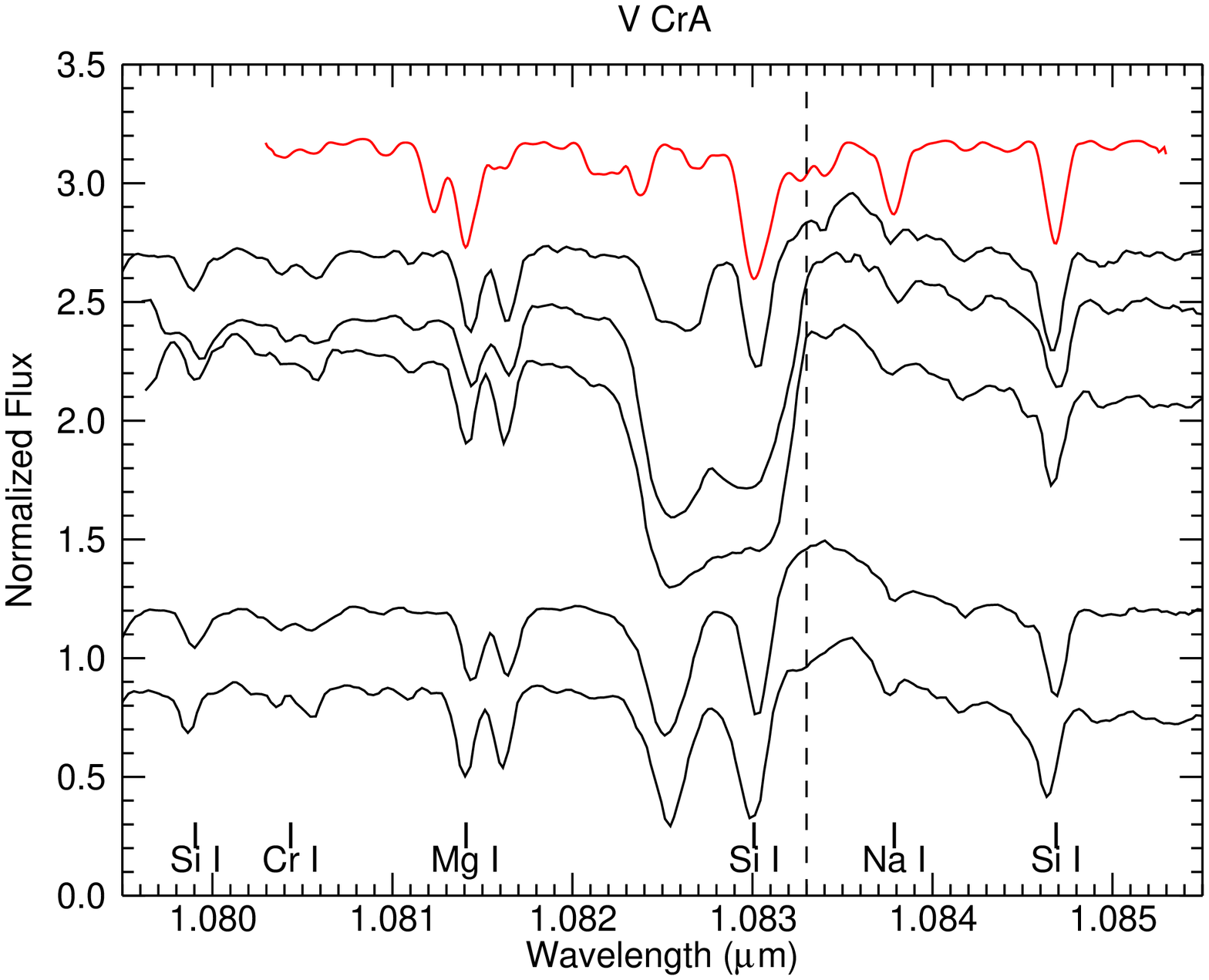}
\hspace{0.5in}
\includegraphics*[width=3.25in]{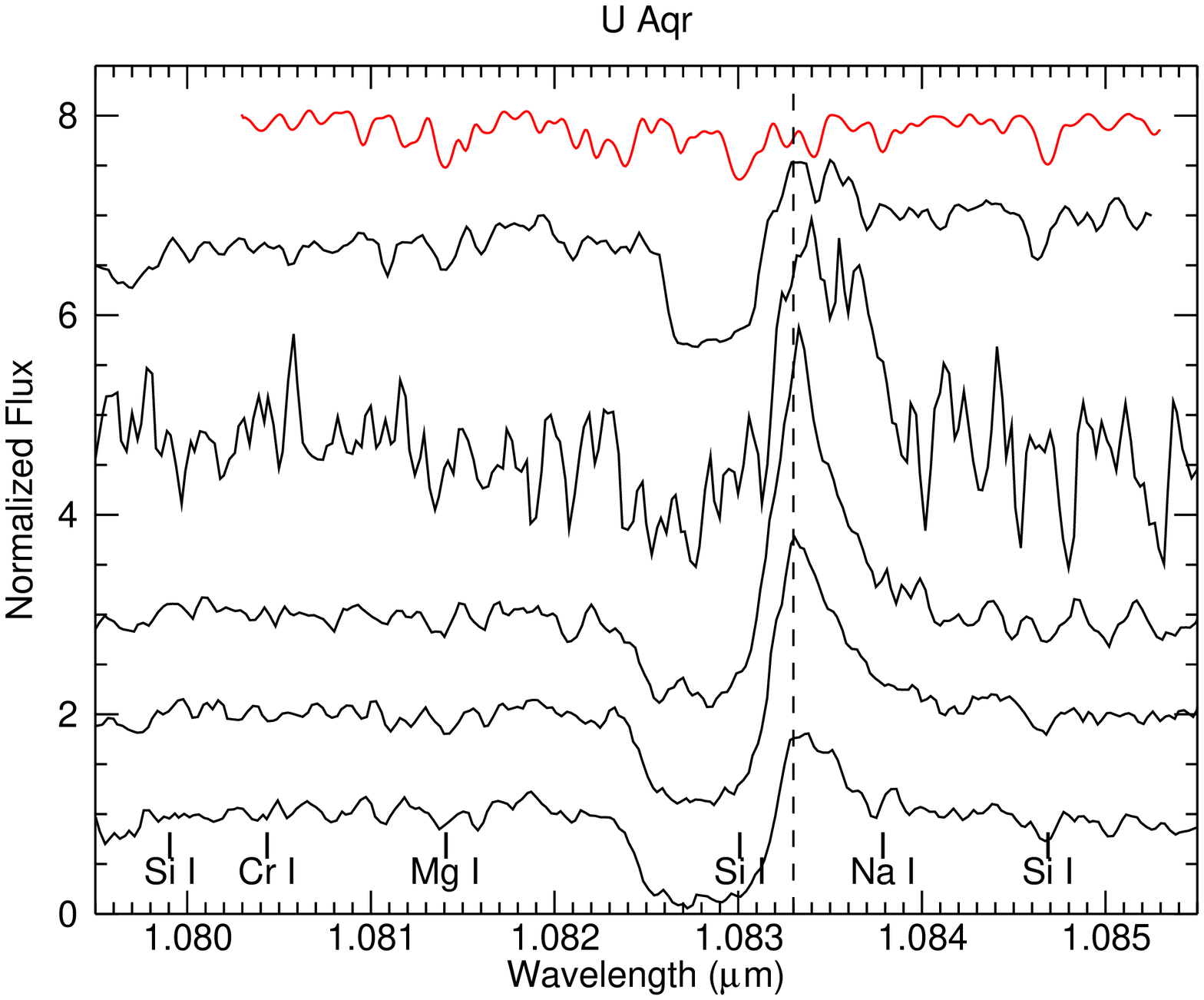}
\includegraphics*[width=3.25in,angle=0]{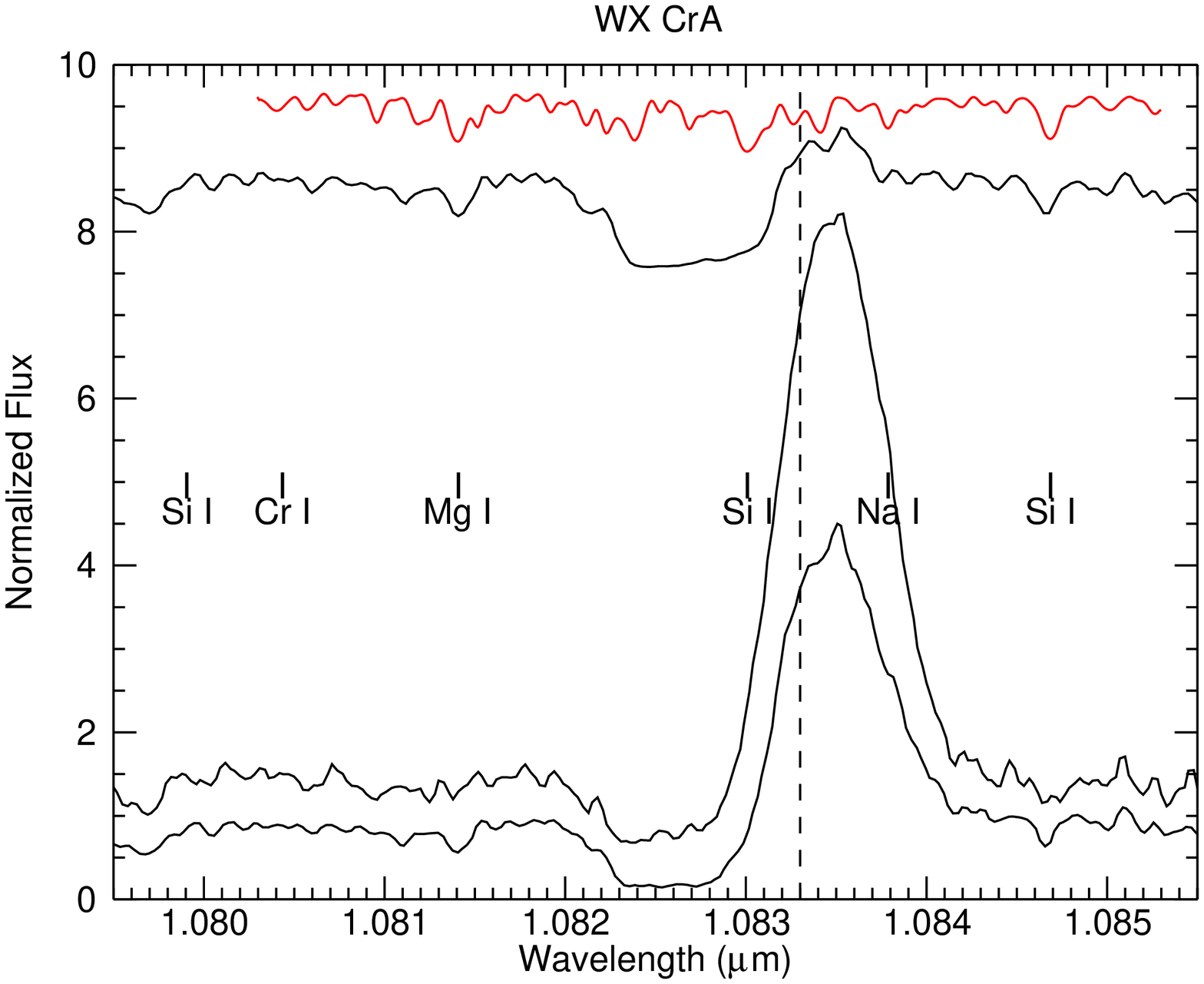}
\hspace{0.5in}
\includegraphics*[width=3.25in]{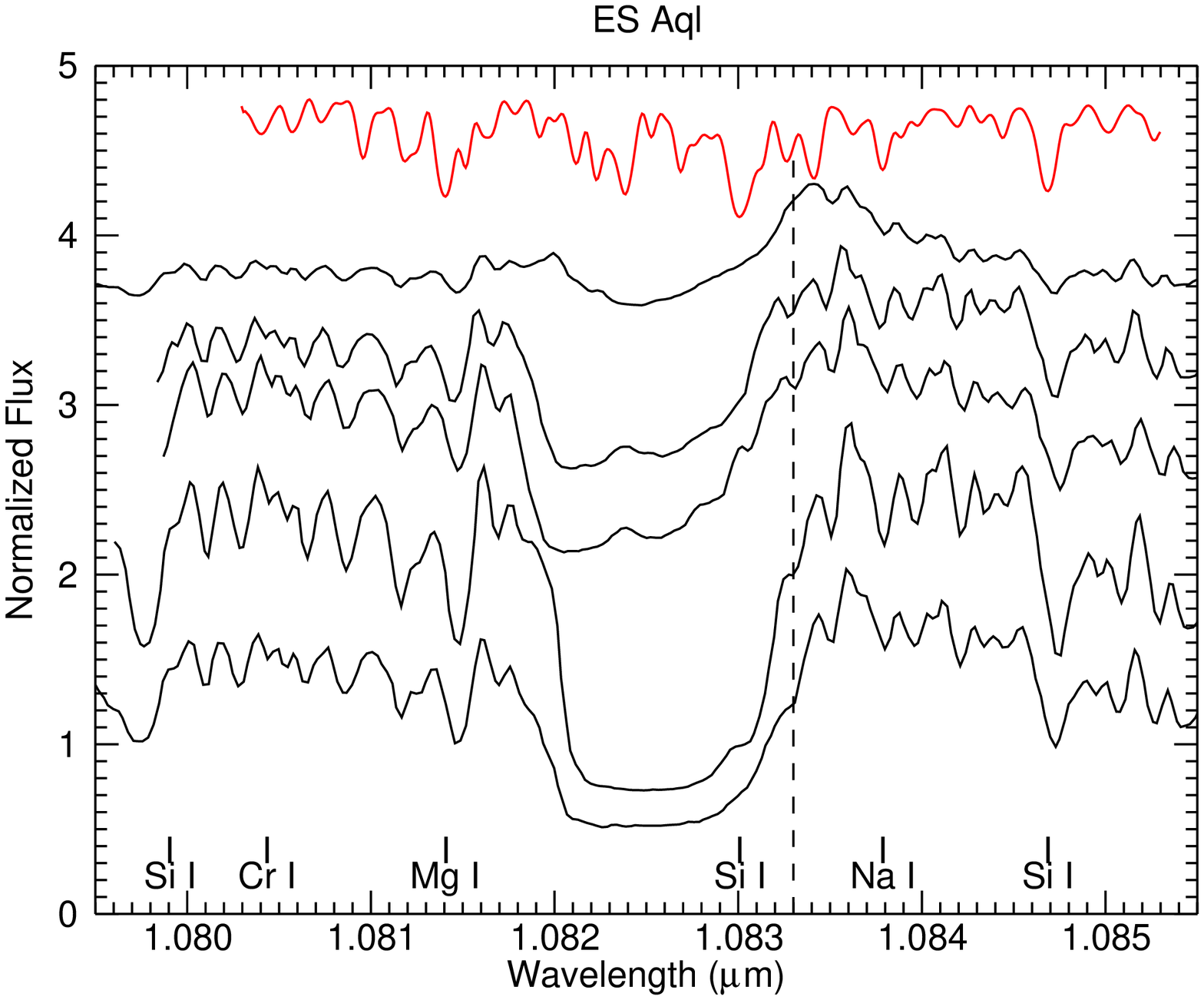}
\includegraphics*[width=3.25in]{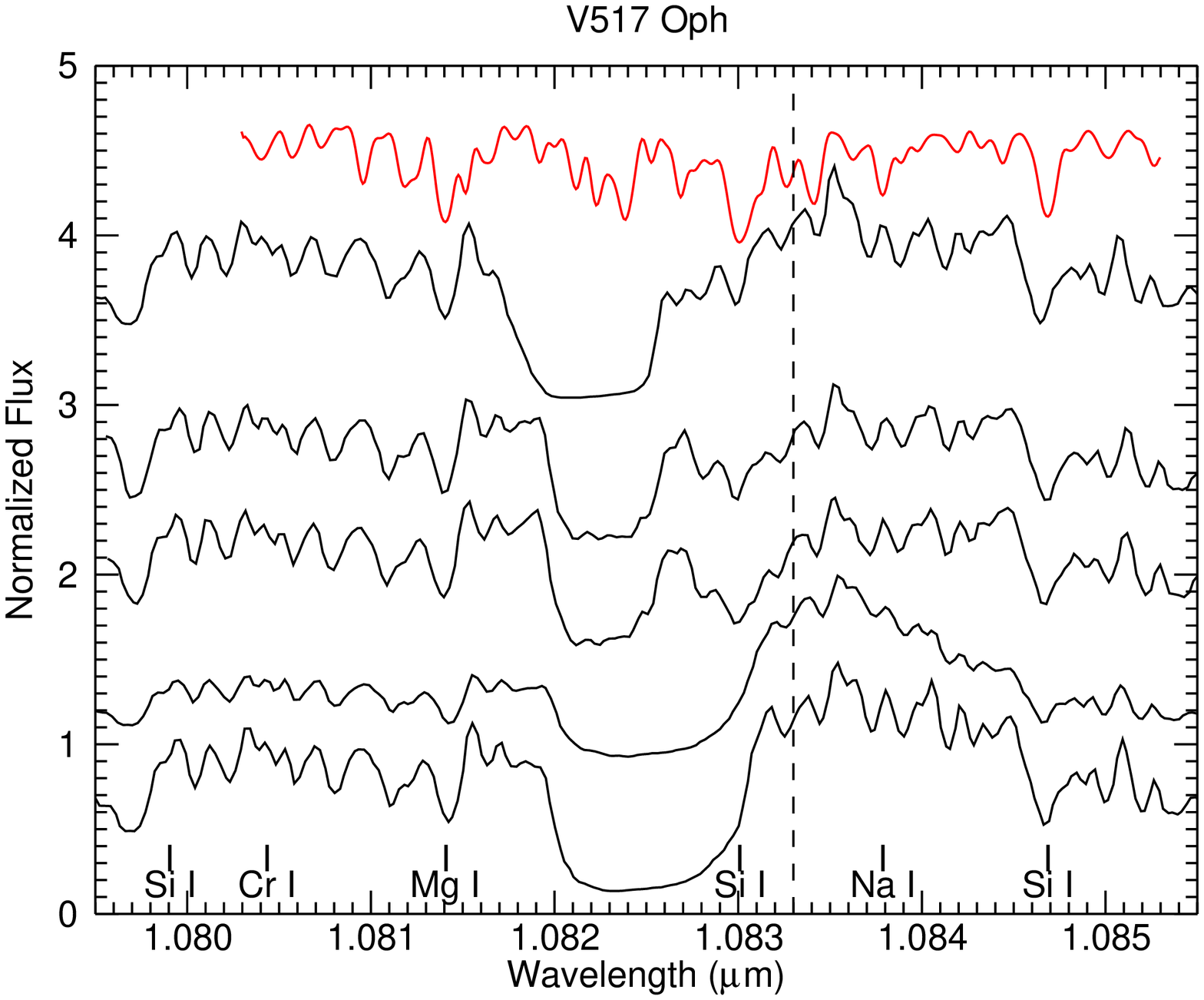}
\hspace{0.5in}
\includegraphics*[width=3.25in]{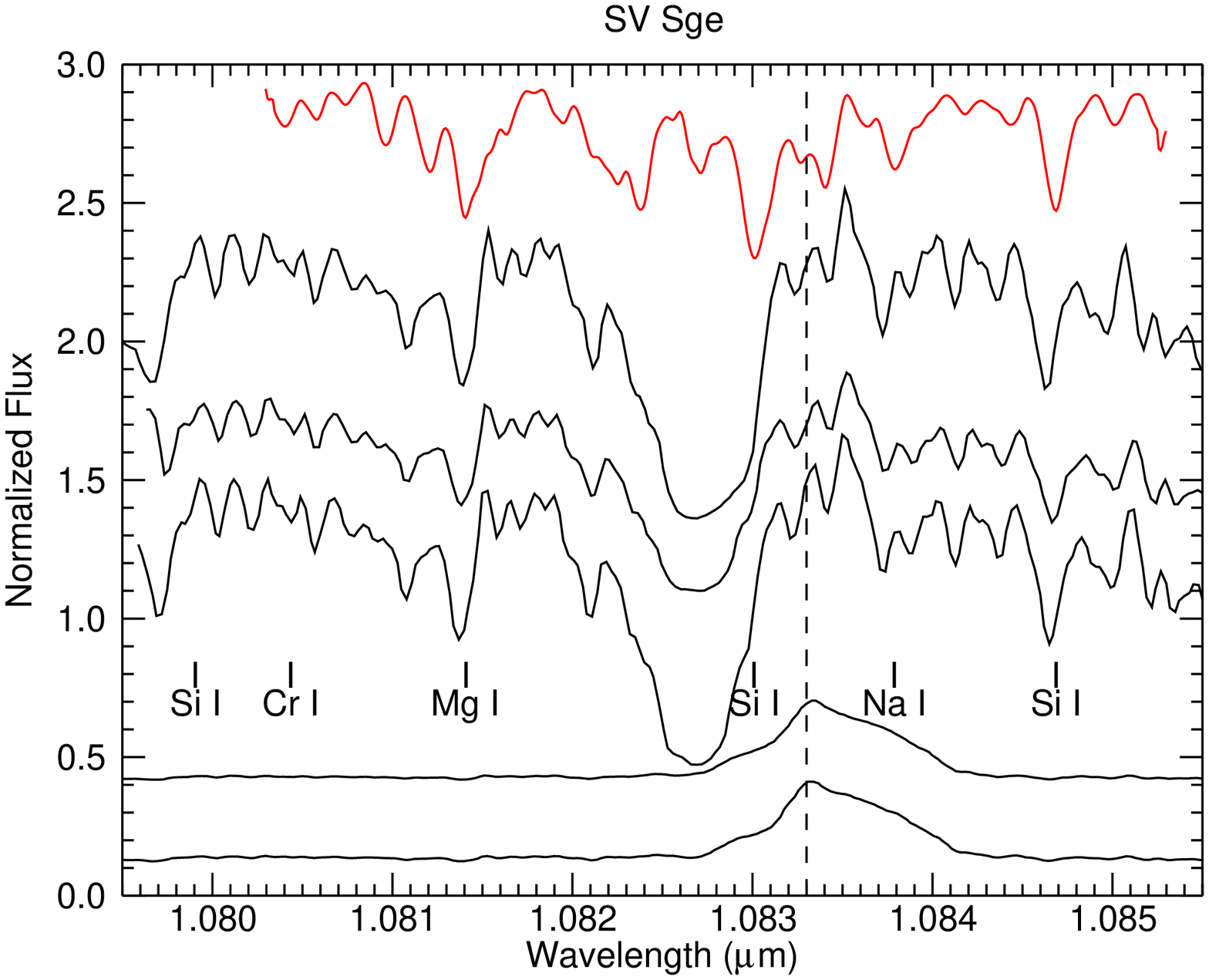}
\caption{Same as Fig. 1a, except that red spectrum is a 5400 K RCB stellar 
model except for V CrA which shows the 6500 K RCB model spectrum.}
\end{figure}

\begin{figure}
\figurenum{2a} 
\includegraphics[width=6in,angle=0]{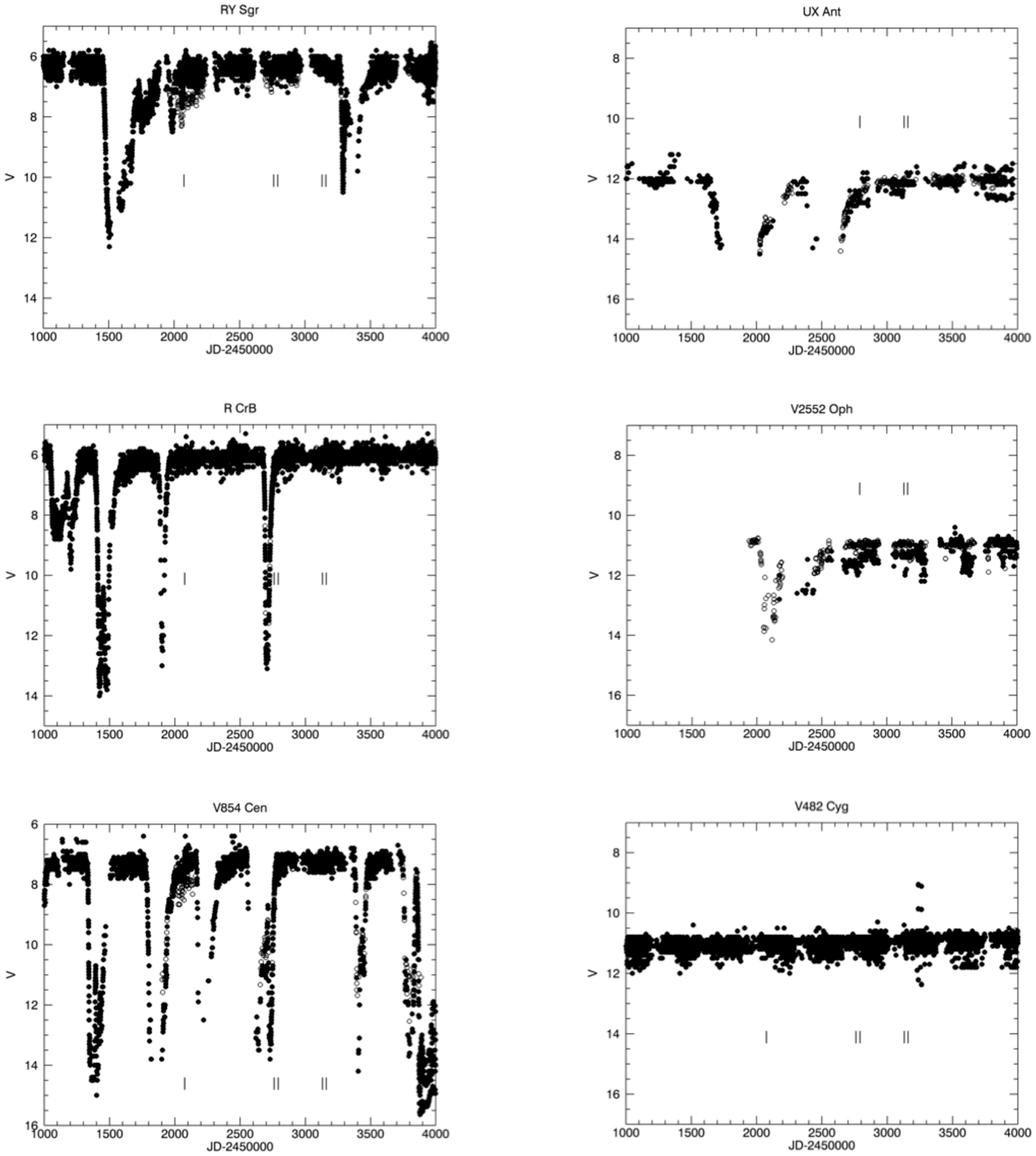} \hspace{0.5in}

\caption{AAVSO (filled circles) and ASAS-3 (open circles) photometry of the sample RCB stars. The epochs of the He I spectra, listed in Table 1, are marked by vertical lines.}
\end{figure}

\begin{figure}
\figurenum{2b} 

\includegraphics*[width=6in,angle=0]{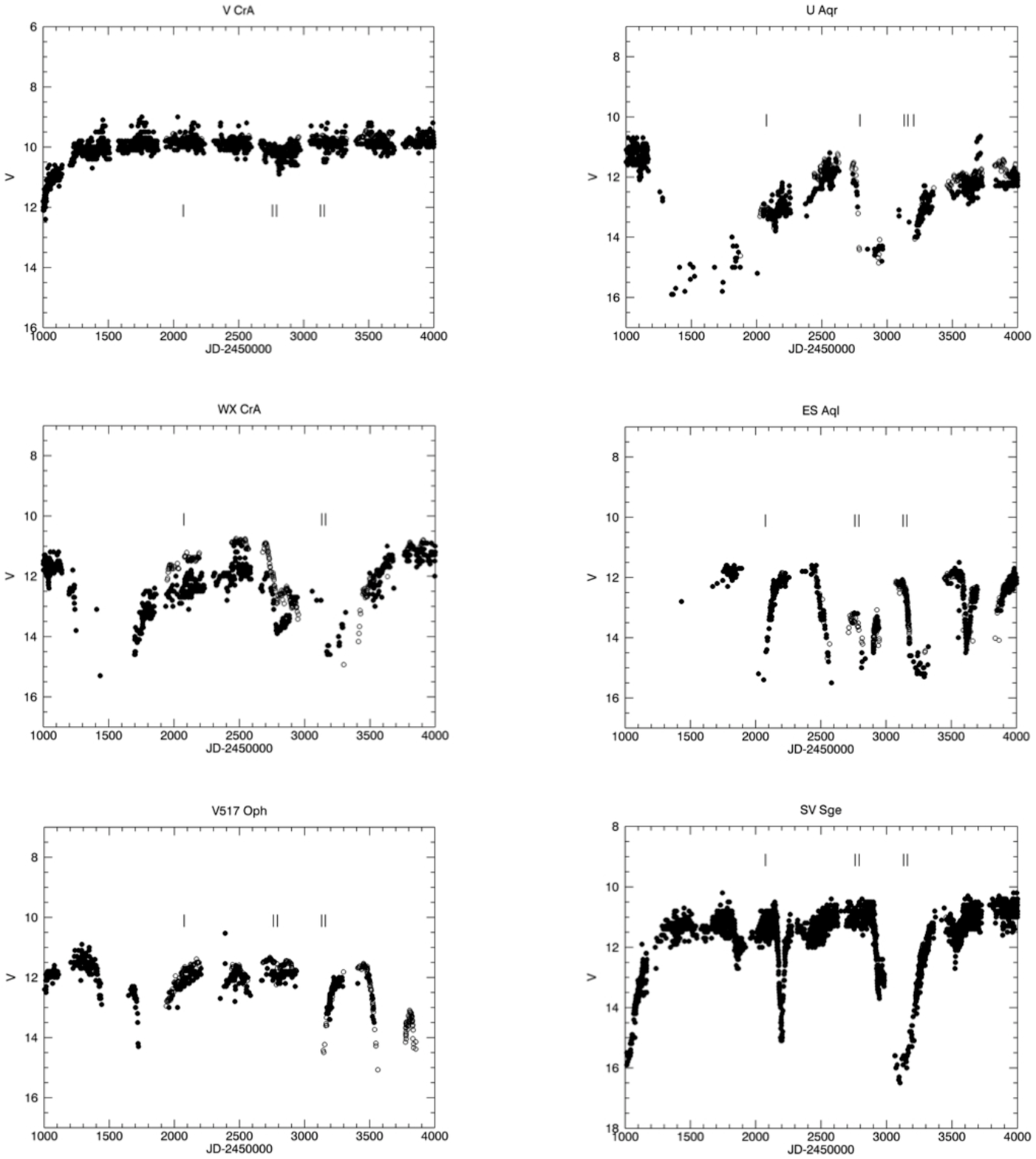}

\caption{AAVSO (filled circles) and ASAS-3 (open circles) photometry of the sample RCB stars. The epochs of the He I spectra, listed in Table 1, are marked by vertical lines.}
\end{figure}

\begin{figure}
\figurenum{3} 
\includegraphics[width=5in,angle=0]{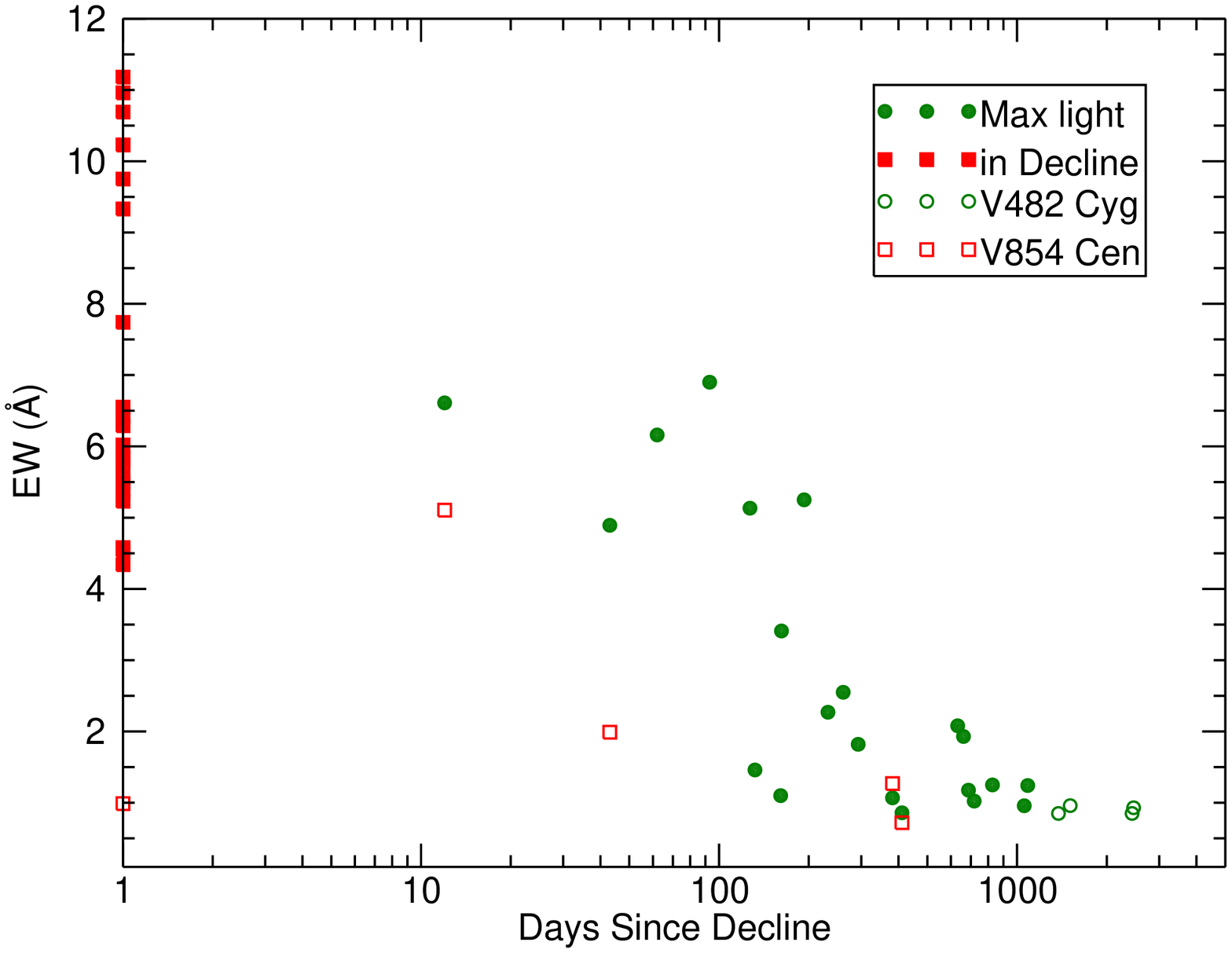}\\
\includegraphics[width=5in,angle=0]{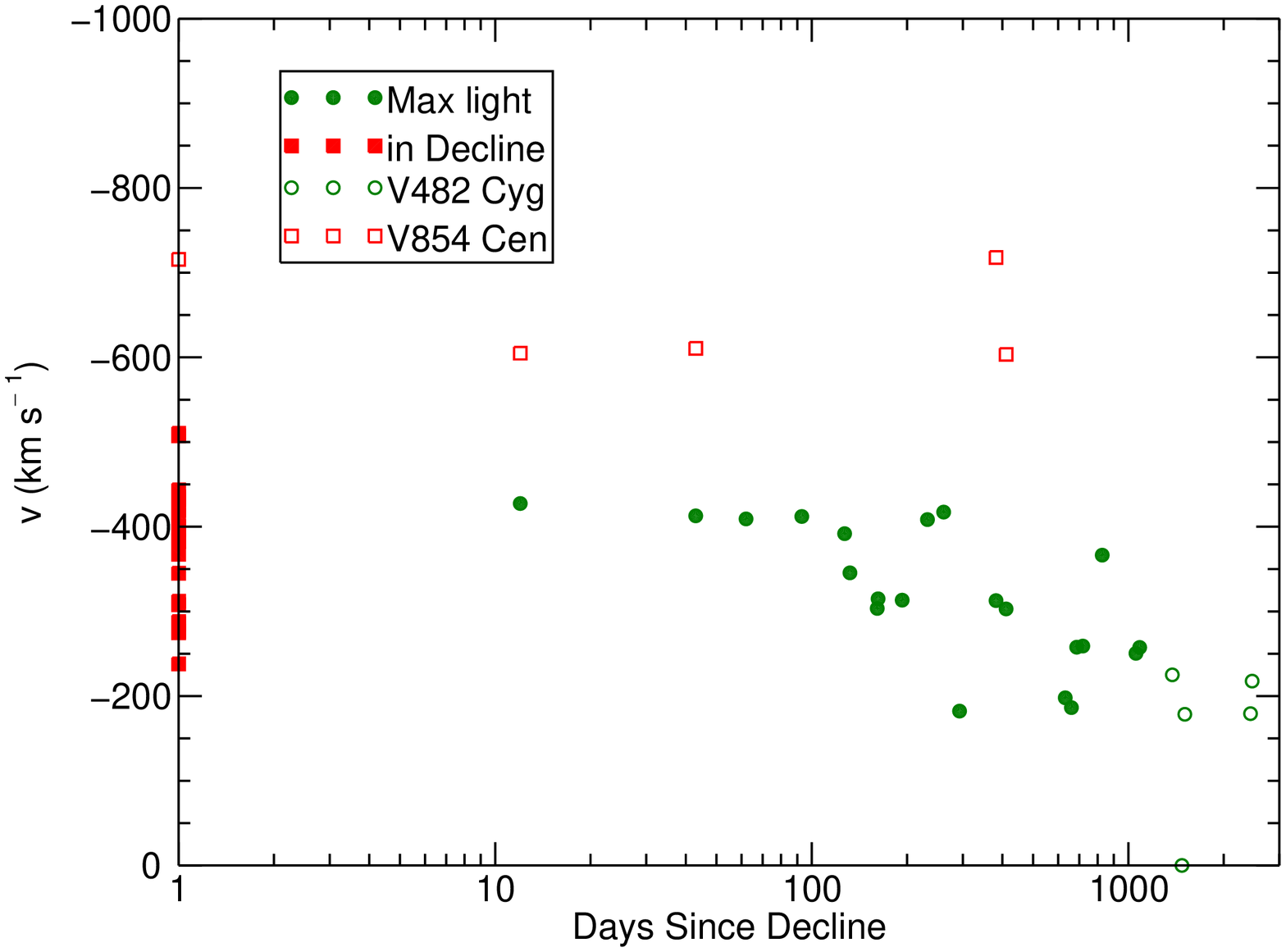}
\caption{Measured equivalent widths (top) and terminal velocities (bottom) of 
the P-Cygni absorption trough, for each epoch of the twelve stars listed 
in Table 1, plotted against time in days since the end of the most 
recent decline. Data points plotted at Day 1 are values observed during 
declines. Filled symbols are measurements for all stars except V482 Cyg 
and V854 Cen.}
\end{figure}


\end{document}